\documentclass[aps,prd,superscriptaddress,showkeys,preprint,nofootinbib]{revtex4}
\usepackage{graphics,epsfig}
\usepackage{subfigure}
\usepackage{float}
\usepackage{epstopdf}
\usepackage{graphicx}
\usepackage{dcolumn}
\usepackage{amsmath}
\usepackage{epstopdf}
\usepackage{svg}
\usepackage[colorlinks=true, linkcolor=blue, citecolor=blue, urlcolor=blue]{hyperref}
\newcommand{\orcid}[1]{\href{https://orcid.org/#1}{\includegraphics[width=8pt]{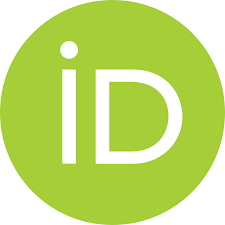}}}
\usepackage{nameref}

\begin{document}
\title{Equatorial light bending around a Hairy Kiselev Black Hole}

\author{Shrishti Kukreti\,\orcid{0000-0006-2033-6515}}
\email{shrishtik.phy30@gmail.com}
\affiliation{Department of Physics, Hemvati Nandan Bahuguna Garhwal Central University, Srinagar-246174, Uttarakhand, India}\vskip 10pt

\author{Shubham Kala\,\orcid{0000-0003-2379-0204}}
\email{shubhamkala871@gmail.com}
\affiliation{The Institute of Mathematical Sciences, C.I.T Campus, Taramani-600113, Chennai, Tamil Nadu, India }

\author{Hemwati Nandan\,\orcid{0000-0002-1183-4727}}
\email{hnandan@associates.iucaa.in}
\affiliation{Department of Physics, Hemvati Nandan Bahuguna Garhwal Central University, Srinagar-246174, Uttarakhand, India}

\author{Faizuddin Ahmed\,\orcid{0000-0003-2196-9622}}
\email{faizuddinahmed15@gmail.com}
\affiliation{Department of Physics, Royal Global University, Guwahati-781035, Assam, India}

\author{Saswati Roy\,\orcid{0000-0002-7028-2627}}
\email{sr.phy2011@yahoo.com}
\affiliation{Department of Physics, National Institute of Technology, Agartala-799046, Tripura, India}
\begin{abstract}
We investigate the deflection angle of light rays confined to the equatorial plane of a Hairy Kiselev black hole. The analysis includes a thorough study of the horizon structure and critical parameters, leading to an analytic expression for the deflection angle in terms of elliptic integrals. Our results confirm that the deflection angle decreases with increasing impact parameter, in agreement with classical predictions of gravitational lensing. The influence of the scalar field, characterized by the coupling constant, shows a nontrivial effect: while moderate values of the coupling constant initially enhance light bending, further increases lead to a suppression of the deflection in the strong-field regime due to modifications in spacetime geometry. Comparative analysis among the Schwarzschild, Kiselev, and Hairy Kiselev black holes indicates that the presence of a quintessential field tends to enhance the deflection, whereas the scalar hair component reduces it. These findings underscore the significant role of scalar fields and exotic matter distributions in shaping light propagation in a modified gravity scenario.
\end{abstract} 


\maketitle

\section{Introduction} \label{Sec1}
\noindent Recent observational breakthroughs, such as the imaging of black hole (BH) shadows by the Event Horizon Telescope (EHT), have provided compelling and direct evidence for the existence of BHs \cite{EventHorizonTelescope:2019dse,EventHorizonTelescope:2019ggy}. These findings have transformed BHs from purely theoretical constructs into confirmed astrophysical entities. Over a century has passed since the Schwarzschild solution to Einstein’s field equations (EFE) first predicted the existence of these exotic objects within the framework of General Relativity (GR) \cite{Schwarzschild:1916uq}. Since then, BHs have evolved from mere mathematical solutions to fundamental components of our understanding of the universe. Beyond the standard solutions of GR, a wide array of BH models have been proposed within the context of modified and alternative theories of gravity. These include, but are not limited to, scalar-tensor theories, Einstein-Gauss-Bonnet (EGB) gravity, f(R) gravity, and massive gravity frameworks \cite{Bronnikov:1973fh,Maartens:2003tw,DeFelice:2010aj,Harko:2011kv,Capozziello:2011et,Sotiriou:2013qea}. These alternative models aim to address the limitations of GR, such as the nature of dark energy, singularity resolution, or quantum gravity effects \cite{Donoghue:1994dn,Satheeshkumar:2021zvl,Carballo-Rubio:2018jzw,Errahmani:2024ran}.

The first quantitative prediction of light deflection by a gravitational field was made by Johann Georg von Soldner in 1801 within the framework of Newtonian mechanics \cite{soldner1804deflection}. By treating light as a stream of particles, he estimated that a light ray passing near the Sun would experience a deflection of approximately 0.87 arc seconds. Over a century later, Albert Einstein revisited the problem using the principles of GR, which accounts for the curvature of both space and time. His calculations yielded a deflection angle of 1.75 arc seconds, precisely twice the Newtonian prediction. This theoretical advancement was empirically validated during the 1919 total solar eclipse by Sir Arthur Eddington and his team, who measured the apparent shift in the positions of stars near the Sun's limb \cite{Dyson:1920cwa}. Their observations closely matched Einstein’s prediction, providing the first direct experimental confirmation of GR and marking a pivotal moment in the history of modern physics. The earliest analytical investigation of light deflection in the strong-field regime, particularly near the photon sphere, was conducted by Darwin in 1959 \cite{darwin1959gravity}, where he obtained an approximate expression for the bending angle by expanding around the unstable circular photon orbit. Following Darwin’s foundational work, several researchers advanced the theoretical understanding of gravitational light bending in the strong-field regime. Atkinson \cite{atkinson1965light} investigated light propagation near compact objects and explored photon trajectories under extreme gravitational fields. Luminet \cite{luminet1979image} provided one of the first detailed simulations of BH shadows and the visual appearance of accretion disks, incorporating relativistic lensing effects. Chandrasekhar \cite{chandrasekhar1998mathematical}, in his seminal monograph, presented a comprehensive mathematical treatment of photon orbits and deflection angles in Schwarzschild and Kerr geometries. Ohanian \cite{ohanian1987black} refined earlier estimates by offering improved analytical expressions for the deflection angle and clarified the physical interpretation of the bending angle in curved spacetimes.

In the early 2000s, a resurgence of interest in strong gravitational lensing led to significant theoretical advancements. Virbhadra and Ellis \cite{virbhadra2000schwarzschild} introduced a lensing framework for strong deflection by BHs and demonstrated that relativistic images could be formed near the photon sphere. Concurrently, Frittelli et al. \cite{frittelli2000spacetime} provided a rigorous mathematical foundation by deriving lens equations directly from spacetime geometry using the null geodesic structure. Eiroa et al. \cite{eiroa2002reissner}  extended these studies by analyzing strong lensing observables for various BH metrics. Petters \cite{petters2003relativistic} contributed to the mathematical formalism by developing a general theory for gravitational lensing (GL) in the strong-field limit using singularity theory. Perlick \cite{perlick2004exact} focused on exact lens equations and classification of caustics in spherically symmetric and static spacetimes. Bozza and collaborators \cite{bozza2001strong} made extensive contributions by formulating the strong deflection limit (SDL) approach, which provides analytical expressions for deflection angles and observable parameters in various BH backgrounds. In addition, S.V. Iyer and A.O. Petters  applied the SDL formalism to both Schwarzschild and Kerr BHs, deriving explicit expressions for deflection angles and lensing coefficients, thus highlighting spin-induced asymmetry in Kerr lensing \cite{Iyer:2006cn,Iyer:2009wa}. Uniyal et al. \cite{Uniyal:2018ngj} extended this formalism to the Kerr–Sen BH, exploring the effect of dilaton-axion fields on light deflection and providing a deeper understanding of string-theoretic corrections. More recently, Y.W. Hsiao et al. \cite{Hsiao:2019ohy} investigated GL by Kerr–Newman BHs, offering a comprehensive treatment of deflection angles in both weak and strong field limits, and analyzing how charge and spin jointly influence image formation. Numerous studies have been conducted on GL, among which a few representative works have been cited herein \cite{Ishihara:2016vdc,Beloborodov:2002mr,Bjerrum-Bohr:2014zsa,Ovgun:2019wej,Cunha:2019hzj,Shaikh:2019itn,Sotani:2015ewa,Moffat:2008gi,Liu:2015zou,Majumdar:2004mz,Tsukamoto:2014dta,Fu:2021fxn,Heydari-Fard:2021pjc,Kumaran:2023brp,Paul:2020ufc,Lim:2016lqv,Kala:2020prt,Kala:2020viz,Parbin:2023zik,Javed:2023iih,Nazari:2022fbn,Kala:2022uog,Ovgun:2019qzc,Ovgun:2025ctx,Pantig:2022ely,Kuang:2022xjp,Li:2020dln,Abdujabbarov:2017pfw,Atamurotov:2023rye,Kala:2024fvg,Wang:2024iwt,Islam:2022ybr,Kumar:2020hgm,Kumar:2020sag,Vishvakarma:2024icz,Pantig:2024ixc,Kala:2025iri,Roy:2025hdw,Roy:2025qmx}. 

In modern cosmology, the accelerated expansion of the Universe was first confirmed through observations of \textit{Type Ia supernovae} and is attributed to a mysterious component known as dark energy \cite{Riess:1998dv,SupernovaSearchTeam:2004lze}. Within the standard $\Lambda$CDM model, this is typically modeled by a cosmological constant ($\Lambda$) \cite{Peebles:2002gy}. However, dynamical alternatives such as quintessence, a slowly rolling scalar field with an evolving equation of state $w_q \in (-1, -1/3)$, have been proposed to address the fine-tuning and coincidence problems associated with $\Lambda$ \cite{DiPietro:1999ia}. The solution for a spherically symmetric spacetime geometry surrounded by a quintessence matter is first studied by Kiselev \cite{Kiselev_2003}. When BHs are considered in such non-vacuum backgrounds, particularly surrounded by quintessence-like fields, the spacetime geometry is significantly altered from classical solutions like Schwar-zschild or Kerr. These modifications have direct consequences on the propagation of photons, leading to measurable changes in GL observables, such as the bending angle, photon sphere structure, and the location of relativistic images.

GL by BHs in a quintessence-dominated background therefore provides a compelling framework to study deviations from GR and standard cosmology. This becomes particularly relevant in the era of precision observations through instruments like the Event Horizon Telescope (EHT) and future space-based missions like Square Kilometre Array (SKA) and Laser Interferometer Space Antenna (LISA), which are capable of probing the near-horizon structure of compact objects \cite{Harrison:2016stv,McCarty:2024oau,Gao:2021sxw}. Accurate modeling of such systems may reveal subtle imprints of dark energy fields, providing complementary constraints to cosmological measurements from supernovae and the cosmic microwave background (CMB). In this context, analyzing light deflection in BH spacetimes influenced by quintessence not only enriches our understanding of strong-field gravity but also connects astrophysical observations with the underlying dynamics of the Universe’s expansion. The study of GL around BHs in the presence of quintessence has garnered significant interest, and various authors have contributed to this area. Among these, we have cited several relevant works \cite{Younas_2015,ABBAS2021100750,Javed_2020,doi:10.1142/S0217732321502242,kala2025nullgeodesicsthermodynamicsweak,Azreg_A_nou_2017,Mustafa_2022, Finelli:2006iz,Liu:2008sg,Fernando:2014rsa,Atamurotov:2022knb,Molla:2023yxn}. 

The organization of the present paper is as follows: In Section \ref{Sec2}, we provide a detailed description of the Hairy Kiselev BH solution, followed by an analysis of the horizon structure in the subsequent subsection. In Section \ref{Sec3}, we derive the geodesic equations and analyze the corresponding effective potential. Section \ref{Sec4} focuses on the study of critical parameters and the photon trajectory equation. In Section \ref{Sec5}, we obtain an exact expression for the bending angle in terms of elliptic integrals and present a comprehensive graphical analysis. Finally, Section \ref{Sec6} concludes the paper with a summary of the main results and their physical implications.

\section{Hairy Kiselev BH Spacetime} \label{Sec2}
\noindent V.V. Kiselev proposed an exact static spherically symmetric solution to EFEs surrounded by a quintessential field and characterized by the equation of state parameter $(\omega)$ \cite{Kiselev_2003}. Building upon this framework, recent developments have employed the extended gravitational decoupling method  \cite{PhysRevD.95.104019,OVALLE2019213}, to construct more general BH spacetimes that include additional degrees of freedom, often referred to as "gravitational hairs" \cite{OVALLE2021100744}. This method systematically decouples the gravitational sources, allowing for the inclusion of anisotropic sectors and additional scalar, vector, or tensor fields. Using gravitational decoupling method, a modified version of the Kiselev solution has been developed, featuring non-trivial modifications to the metric components while preserving overall spherical symmetry. The resulting geometry admits extra hair parameters and provides a more flexible model to explore deviations from GR. The metric is given by \cite{PhysRevD.108.044073},
\begin{equation}\label{eq_metric}
    ds^2 = -f(r)\,dt^2 + \frac{1}{f(r)}\,dr^2 + r^2\,d\Omega^2,
\end{equation}
where, the metric function is as
\begin{equation}\label{eq_fr}
    f(r) = 1- \frac{2\,M}{r}  - \frac{N}{r^{3\,\omega + 1}} + \alpha\, \exp\left(-\frac{r}{M- \frac{\alpha\,\ell}{2}}\right),
\end{equation}
and \[
d\Omega^2 = d\theta^2 + \sin^2\theta\ d\phi^2.
\]
Here, $M$ is the mass of the BH and is related to the Schwarzschild mass $\mathcal{M}$ as $M = \mathcal{M}+\frac{\alpha\, \ell}{2}$,\, $\alpha$ is the coupling constant, $\ell$ is a constant parameter with length dimensions associated with the primary hair of the BH, $\omega $ is the state parameter that takes different values depending on the nature of the surrounding field and $N$ is a constant related to the strength of the surrounding field.  $\omega$ and $N$ must have different signs in order to respect weak energy conditions.

In order to proceed further, we perform a coordinate transformation to express the metric in advanced Eddington-Finkelstein coordinates\cite{Cavalcanti_2022}. So \( (t, r, \theta, \phi) \rightarrow (v, r, \theta, \phi) \) and $v$ are given by,
\begin{equation}
    v=t-\int\left(1- \frac{2M}{r}  - \frac{N}{r^{3\omega + 1}} + \alpha e^{-\frac{r}{M- \frac{\alpha l}{2}}}\right)^{-1}dr. 
\end{equation}
So, the metric (\ref{eq_metric}) takes the form\\
\begin{equation}
    ds^2 = -\left(1 - \frac{2M}{r} - \frac{N}{r^{3\omega + 1}} \right.
    \left. + \alpha e^{-\frac{r}{M - \frac{\alpha l}{2}}} \right) dv^2 + 2\epsilon \, dv \, dr + r^2 d\Omega^2.
    \label{metric}
\end{equation}
Here, $v$ is the advanced ($\epsilon =+1$) or retarded ($\epsilon =-1$) Eddington time. Throughout this study, we employ the spacetime metric given in Eq.~\eqref{metric} to investigate null geodesic motion and the deflection of photon rays, highlighting how various key parameters influence these phenomena. As an illustrative example, we choose the state parameter \(w = -2/3\) and discuss the resulting outcomes, noting that other values of \(w\) can be treated in a similar manner.
\subsection{Horizon Structure}
\begin{figure} [H]
	\begin{center}
    \begin{subfigure}[]
     {\includegraphics[width=0.45\textwidth,height=0.3\textheight]{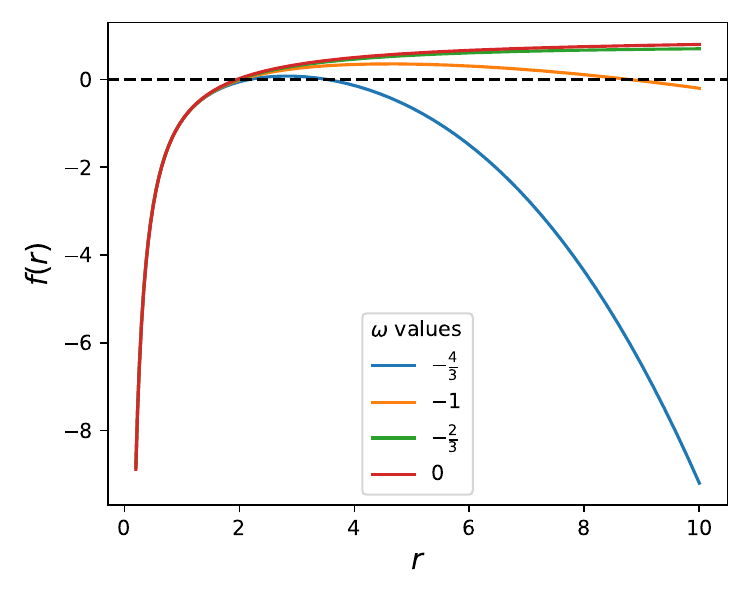}}   
    \end{subfigure}
     \begin{subfigure}[]
     {\includegraphics[width=0.45\textwidth,height=0.3\textheight]{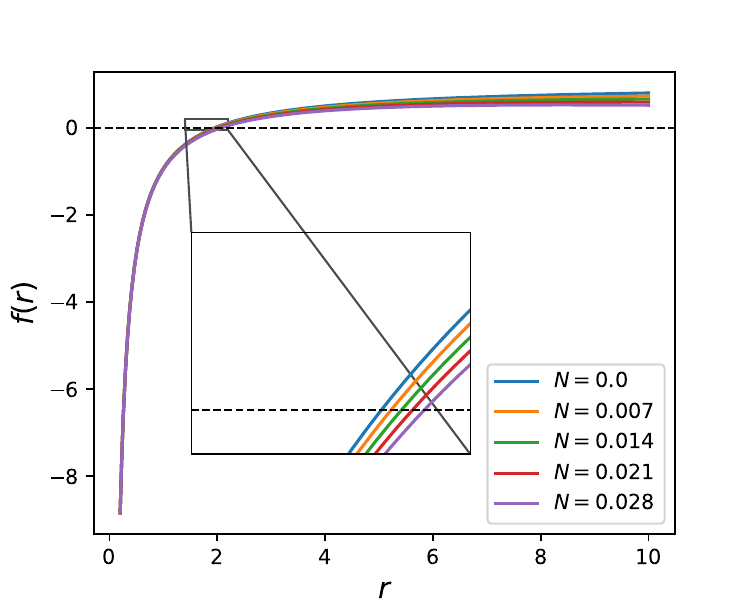}}  
    \end{subfigure} 
    \begin{subfigure}[]
     {\includegraphics[width=0.45\textwidth,height=0.3\textheight]{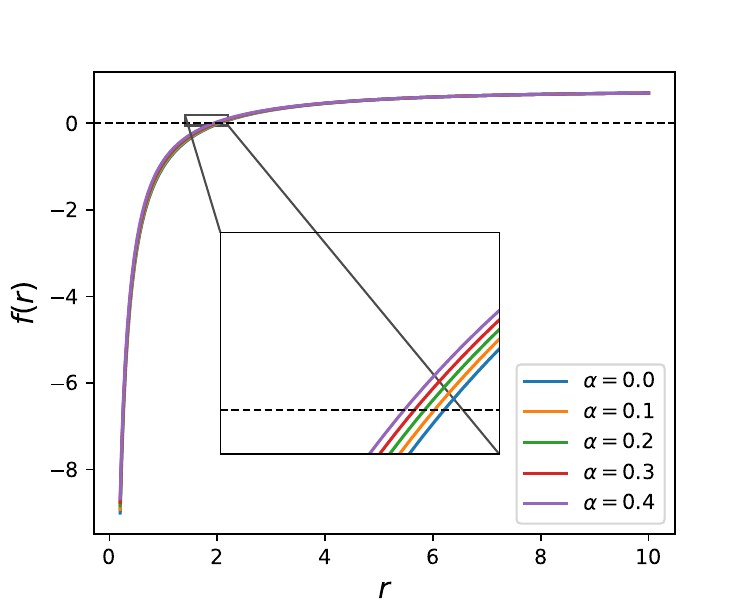}}  
    \end{subfigure} 
	\end{center}
	\caption{The variation of metric function $f(r)$ with radial distance. (a) For different values of state parameter ($\omega$), here we consider, $M=1$, $N=0.007$, $\alpha=0.2$ and $\ell=0.05$. (b) For different values of $N$, here we consider $M=1$, $\alpha=0.2$, $\omega=-2/3$ and $\ell=0.05$. (c) For different values of $\alpha$, here we consider $M=1$, $N=0.007$, $\omega=-2/3$ and $\ell=0.05$.} \label{f(r)variation}
\end{figure}

\noindent In GR, the location of a BH horizon is determined by the condition under which the spacetime metric component $g_{tt}$(or equivalently the lapse function $f(r)$) vanishes. This corresponds to the radius at which outgoing light rays become trapped, marking the boundary of the causal region, the event horizon. For static and spherically symmetric spacetimes, such horizons are found by solving the equation where the metric function equals zero. In Fig. \ref{f(r)variation},  we present the variation of the metric function $f(r)$ with respect to the radial coordinate $r$ for different values of the state parameter $(\omega)$, the surrounding field parameter $(N)$, and the coupling constant $(\alpha)$. The graphical analysis reveals that for $\omega = -\frac{4}{3}$ and $\omega = -1$, the metric function possesses two real and positive roots, indicating the existence of two horizons. This behavior is consistent with BH spacetimes surrounded by exotic fields capable of generating repulsive effects at large distances. On the other hand, for $\omega = -\frac{2}{3}$ and $\omega = 0$, the metric function intersects the radial-axis only once, signifying the presence of a single BH horizon. No other horizon is observed in these cases, which implies that the repulsive effects are insufficient to produce an additional outer horizon. To examine this behavior further, we fix the equation of state parameter at $\omega = -\frac{2}{3}$ and analyze the influence of varying $N$ and $\alpha$ on the horizon structure. The results show that even with different values of $N$ and $\alpha$, the number of horizons remains unchanged; only a single horizon forms. This indicates that for a fixed $\omega = -\frac{2}{3}$, neither the coupling strength nor the density of the surrounding field are sufficient to modify the qualitative nature of the spacetime geometry. Therefore, the BH retains a single horizon irrespective of changes in $N$ and $\alpha$ within the considered range.

\begin{figure} [H]
	\begin{center}
    \begin{subfigure}[]
     {\includegraphics[width=0.6\textwidth,height=0.35\textheight]{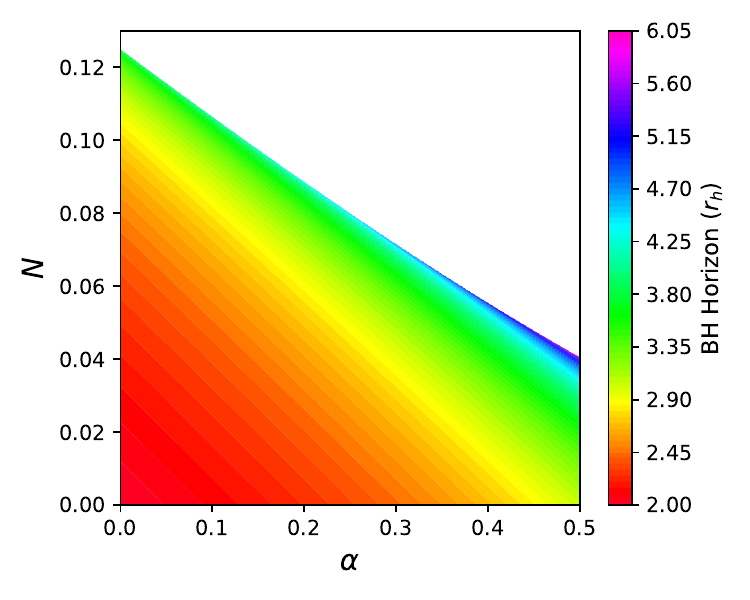}}   
    \end{subfigure}
	\end{center}
	\caption{Variation of BH horizon with $N$ and $\alpha$. Here, we consider $M=1$, $\omega=-2/3$ and $\ell=0.05$.} \label{HorzDens}
\end{figure} 
Here, we focus on a particularly interesting case, namely $\omega = -\frac{2}{3}$, for which an analytical expression for the BH horizon becomes tractable. In this scenario, the spacetime structure is simplified, allowing the metric function to yield a closed-form solution for the horizon radius. Thus, the only physically valid BH horizon in this case is given by
\begin{equation}
    r_h = \frac{(1 - \alpha \gamma) - \sqrt{(\alpha \gamma - 1)^2 - 4 N (2M - \alpha)}}{2N},
\end{equation}
with $\gamma = \frac{2}{2M - \alpha \ell}$.
This expression yields real and positive roots only when $\alpha \gamma > 1 + \sqrt{4N(2M-N \alpha)}$. The variation of the BH horizon radius with respect to the parameters $N$ and $\alpha$ is illustrated in Fig.~\ref{HorzDens}. It is clearly observed that the horizon radius increases with increasing values of both parameters. This behavior highlights the influence of the surrounding field parameter $(N)$ and the coupling constant $(\alpha)$ on the causal structure of spacetime. A larger horizon radius implies a stronger deviation from the standard Schwarzschild geometry. Notably, in the limiting case where $N \rightarrow 0$ and $\alpha = 0$, the solution consistently reduces to the well-known Schwarzschild BH in GR, with a horizon located at $r_h = 2(M=1)$. 
\section{Geodesic Equations and Effective Potential}\label{Sec3}
\noindent In the context of BH physics, null geodesics constitute an essential framework for investigating the causal structure and curvature properties of spacetime, as they govern the trajectories of massless particles such as photons. They play a crucial role in understanding several important phenomena such as the photon sphere, the stability of circular orbits, and the deflection of light (photon trajectories) in the vicinity of a BH. These features are not only essential for theoretical insights but also have observable consequences, such as GL and BH shadow formation.

By studying the behavior of null geodesics, one can investigate how various physical parameters affect these geodesic properties. In particular, the influence of various geometric parameters on the location and stability of the photon sphere, the bending angle of light, and the structure of possible photon orbits provides valuable information about the nature of the underlying spacetime and deviations from the standard Schwarzschild black hole solution. Several recent studies on geodesic motion in various black hole solutions have been reported in \cite{Ahmed:2024qeu,Al-Badawi:2025rcq,Ahmed:2025fwz,Al-Badawi:2025uxp,Ahmed:2025oaw,Ahmed:2025sav,Ahmed:2025evv,Al-Badawi:2025rya,Ahmed:2025iqz} and the references therein.
Considering the geodesic motion in a spherically symmetric spacetime, without loss of generality, one can consider the equatorial plane $\theta=\pi/2$. The Lagrangian for a photon travelling in Hairy Kiselev spacetime is given by
\begin{equation}
    2 \mathcal{L} = -f(r)\,\dot{v}^{2} + 2\, \epsilon\, \dot{v}\, \dot{r} + r^{2}\, \dot{\phi}^{2}.
\end{equation}
where the dot sign means the derivative with respect to the proper time $\tau$.
For null geodesics, $2 \mathcal{L}=0$ and from conserved quantities, we obtain
\begin{equation}
    \dot{\phi} = \frac{L}{r^{2}},
\end{equation}
and
\begin{equation}
    \dot{v} = \frac{\epsilon\, \dot{r}+E}{f(r)}.
\end{equation}
Now applying the constraint equation, we obtain the geodesic equation for the radial coordinate $r$ as
\begin{equation}
    -f(r) \left( \frac{\epsilon \dot{r}+E}{f(r)} \right)^{2} + 2 \epsilon \left( \frac{\epsilon \dot{r}+E}{f(r)} \right) \dot{r} + \frac{L^{2}}{r^{2}}=0.
\end{equation}
Subsequently, we can write,
\begin{equation}
    \dot{r}^{2} = E^{2} - f(r)\,\frac{L^{2}}{r^{2}}
\end{equation}
Comparing the above equation with $ \dot{r}^{2} = E^{2} - V_\text{eff}(r)$, we find the effective potential of the null geodesics given by
\begin{equation}
    V_\text{eff}(r) = f(r)\,\frac{L^{2}}{r^{2}}.
\end{equation}
or we can write
\begin{equation}
    V_\text{eff}(r) = \left( 1- \frac{2M}{r}  - \frac{N}{r^{3\omega + 1}} + \alpha e^{-\frac{2r}{2M-\alpha l}}\right) \frac{L^2}{r^2}.\label{potential}
\end{equation}
\begin{figure}[H] 
\includegraphics[width=0.48\textwidth,height=0.3\textheight]{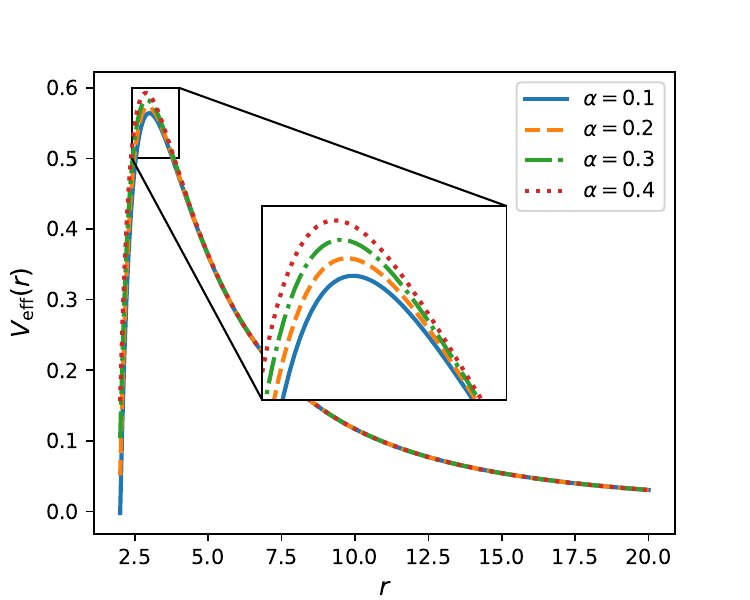}\quad \quad
\includegraphics[width=0.45\textwidth,height=0.28\textheight]{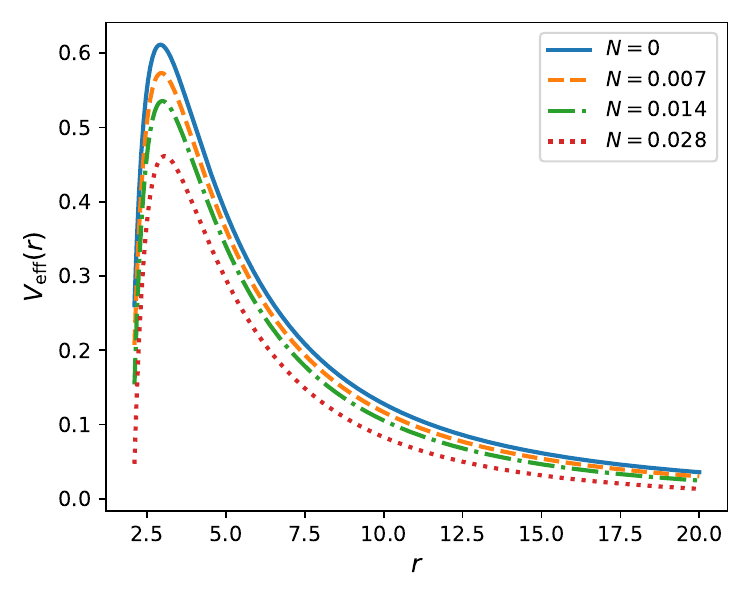}
\caption{ Behavior of the effective potential as a function of radial distance $r$ is shown for varying values of the coupling parameter $\alpha$ (left panel, $N=0.007$) and the normalization constant $N$ of the field (right panel, $\alpha=0.2$). Here, we set the angular momentum $L=4$, the BH mass $M=1$, the constant parameter $\ell=0.05$ and state parameter $\omega=-2/3$.} \label{fig:potential}
\end{figure}
\noindent From the above equation, it becomes evident that the effective potential for null geodesics is influenced by several geometric key parameters. These include the BH mass $M$, the scalar field coupling (or hairy) parameter $\alpha$, the constant parameter $\ell$, the field parameter $N$, and the state parameter $\omega$. Moreover, the angular momentum \(L\) alters the effective potential of photon particles. For an example, setting the state parameter $w=-2/3$, the effective potential for null geodesics from Eq. (\ref{potential}) reduces as
\begin{equation}
    V_\text{eff}(r) = \left( 1- \frac{2M}{r}  -N\,r+ \alpha e^{-\frac{2r}{2M-\alpha l}}\right) \frac{L^2}{r^2}.\label{potential2}
\end{equation}
\noindent
Fig.~\ref{fig:potential} illustrates the variation of the effective potential as a function of the radial coordinate for different values of parameters $\alpha$ and $N$. We observe that increasing the parameter $\alpha$ leads to a higher peak in the effective potential, indicating a stronger gravitational influence or modification due to $\alpha$. In contrast, an increase in the parameter $N$ results in a suppression of the effective potential, suggesting that the spacetime geometry becomes less confining for photons or test particles. Furthermore, the effective potential exhibits only a single maximum and no minimum, which implies the absence of stable circular orbits. The presence of a maximum alone corresponds to an unstable circular orbit, typically associated with the location of the photon sphere.

\section{Critical Variables and the Equation of Path for Photons} \label{Sec4}
\begin{figure} [H]
	\begin{center}
    \begin{subfigure}[]
     {\includegraphics[width=0.6\textwidth,height=0.35\textheight]{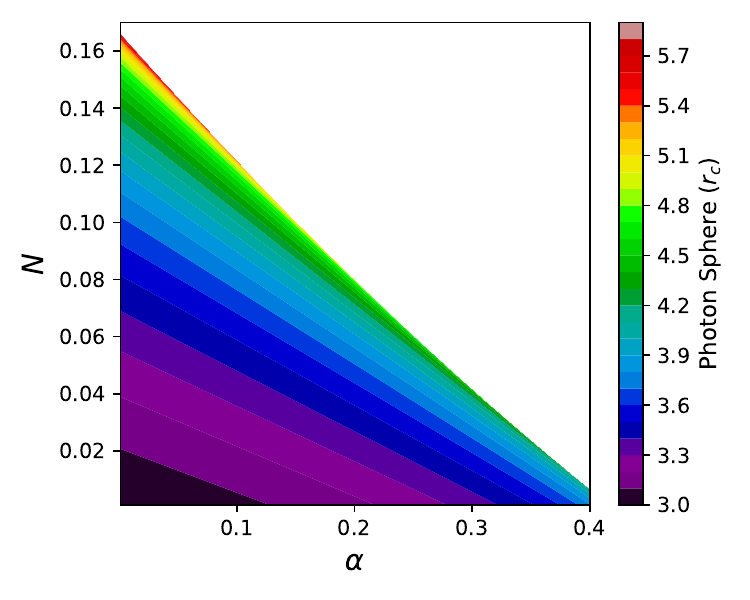}}   
    \end{subfigure}
	\end{center}
	\caption{Variation of BH photon sphere with $N$ and $\alpha$. Here, we consider $M=1$, $\omega=-2/3$ and $\ell=0.05$.} \label{PhotonDens}
\end{figure}

\noindent In this section, we study geometric properties of the metric, specifically focusing on the photon sphere and the trajectory of photon rays, and discuss how geometric key parameters influence these properties. For circular orbits of radius $r=r_c$, conditions $\dot{r}=0$ and $\ddot{r}=0$ must be satisfied. The first condition gives us the critical impact parameter for photon particles. The second condition implies that $V'_\text{eff}(r)=0$ gives us the radius of the photon sphere. In our case, we find the following expression using the potential (\ref{potential2}) in the condition $V'_\text{eff}(r)=0$ as,
\begin{equation}
\frac{6\,M}{r} + N r + 2\,\alpha\, e^{-\frac{2\,r}{2\,M - \alpha\, \ell}} \left( \frac{r}{2\,M - \alpha\,\ell} - 1 \right) = 2
\end{equation}
For small $r$ values, that is, $r<(M-\alpha\,\ell/2)$, the above equation reduces as,
\begin{equation}
\frac{6\,M}{r} + \left(N+ \frac{6\,\alpha}{2\,M - \alpha\, \ell}\right)\,r-2\,(1+\alpha)=0.
\end{equation}
Solving for $r$ gives
\begin{equation}
    r_{c\pm} = \frac{(\alpha\,+1) \pm \sqrt{(\alpha+1)^{2} - 6\,M\,(\alpha\, \beta\, +N)}}{(\alpha\,\beta+N)},
\end{equation}
where $\beta=\frac{6}{2\,M-\alpha\,\ell}$. Here, both \( r_{c+} \) and \( r_{c-} \) lie outside the event horizon \( r_h \), with \( r_{c+} > r_{c-} > r_h \). The larger root \( r_{c+} \) always lies further away from the BH and does not correspond to an unstable null orbit, making it physically irrelevant for photon dynamics. \( r_{c-} \) represents the radius of the unstable circular photon orbit, provided that it remains outside the horizon. Therefore, the radius of an unstable circular orbit for a photon is \( r_{c-} = r_{\text{ps}} \), also called the \textit{photon sphere}. The photon sphere is plotted in Fig.~\ref{PhotonDens} to examine the behavior of the parameters $\alpha$ and $N$. It is clearly observed that the radius of the photon sphere increases with increasing values of both parameters. Furthermore, in the limiting case $\alpha = 0$ and $N = 0$, the result reduces to the standard Schwarzschild BH photon sphere at $r_{ps} = 3(M=1)$, as expected.

We convert the equation of motion in terms of $u=1/r$ and obtain the equation of path for photons as follows,
\begin{equation} \label{eqopath}
  \left(   \frac{du}{d\phi} \right)^{2}- \mathcal{B}(u)=0, 
\end{equation}
where,
\begin{equation} \label{meomp}
\mathcal{B}(u) = \frac{1}{b^{2}} - \left[ 1-2Mu-Nu^{3\omega+1}+\alpha e^{\frac{-2}{u(2M-\alpha l)}} \right]\,u^2.
\end{equation}
Here $b=L/E$ is the impact parameter for the photon ray.

For the critical value of the closest approach, we put $ \frac{du}{d\phi}=0$. Identifying this point of the closest approach as $u=u_{2}$, from Eq. \ref{eqopath}, we have
\begin{equation}
    \frac{1}{b^{2}} = u_{2}^{2}-2Mu_{2}^{3}-Nu_{2}^{3\omega+3} + \alpha u_{2}^2 e^{\frac{-2}{u_{2}(2M-\alpha l)}}.
\end{equation}
Substituting $u_{2}=1/r_{ps}$, we obtain the critical value of impact parameter for circular orbits,
\begin{equation} \label{criticalimpact}
    b_{sc} = \sqrt{\frac{r_{ps}^{3\omega+3}}{r_{ps}^{3\omega+1}-2M r_{ps}^{3\omega}-N+\alpha r_{ps}^{3\omega+1} e^{\frac{-2r_{ps}}{(2M-\alpha l)}}}}.
\end{equation}
Setting state parameter $w=-2/3$, for example, we find
\begin{equation} \label{criticalimpact2}
    b_{sc} = \sqrt{\frac{r^3_{ps}}{r_{ps}-2\,M-N\,r^2_{ps}+\alpha\, r_{ps}\,e^{\frac{-2r_{ps}}{(2M-\alpha l)}}}}.
\end{equation}

According to the circular orbit condition ($B(u)=0$) and solving Eq. \ref{meomp} by setting $w=-2/3$, we get one real root $u_{1}$ and two additional roots $u_{2}$ and $u_{3}$, ($u_{3} > u_{2} > u_{1}$) which are,
 \begin{equation}
    u_{1} = \frac{1-\frac{2M}{r_{0}} -N -\eta - \sqrt{\left(\frac{2M}{r_{0}}+N+\eta-1\right)^{2}-8M \left(\frac{2M+\eta}{r_{0}^{2}}+\frac{N-1}{r_{0}}\right)}}{4M},
\end{equation}
\begin{equation}
    u_{2} = \frac{1}{r_{0}},
\end{equation}
\begin{equation}
    u_{3} = \frac{1-\frac{2M}{r_{0}} -N -\eta + \sqrt{\left(\frac{2M}{r_{0}}+N+\eta-1\right)^{2}-8M \left(\frac{2M+\eta}{r_{0}^{2}}+\frac{N-1}{r_{0}}\right)}}{4M},
\end{equation}

\begin{figure} [H]
	\begin{center}
    \begin{subfigure}[]
     {\includegraphics[width=0.45\textwidth,height=0.3\textheight]{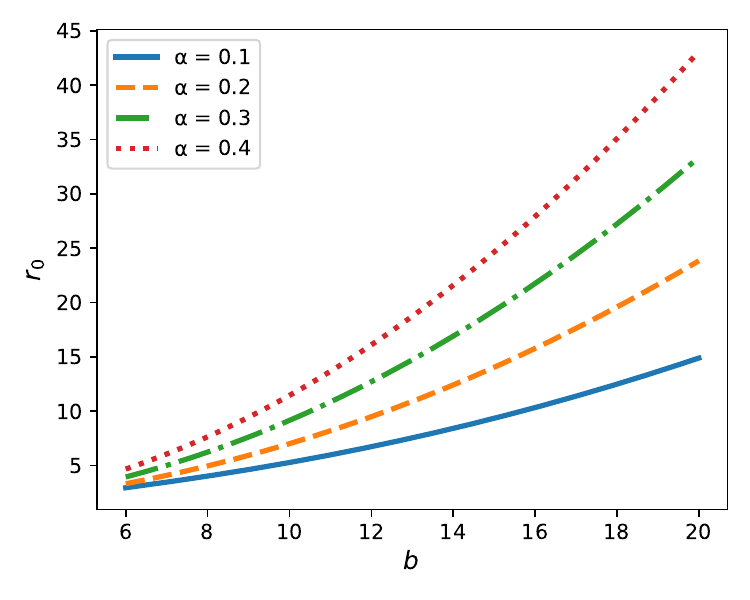}}   
    \end{subfigure}
     \begin{subfigure}[]
     {\includegraphics[width=0.45\textwidth,height=0.3\textheight]{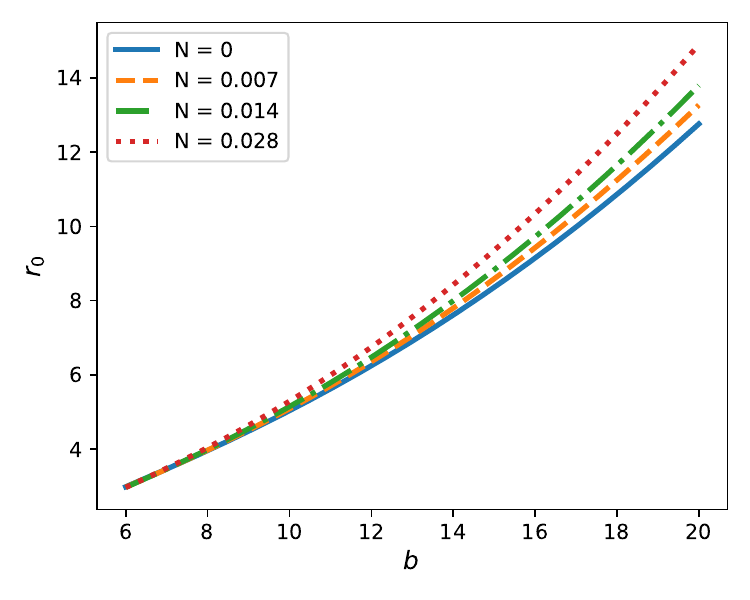}}  
    \end{subfigure}   
	\end{center}
	\caption{Closest approach as as a function of impact parameter $b$ is shown for varying values of the coupling parameter $\alpha$ (left panel, $N=0.007$) and the normalization constant $N$ of the field (right panel, $\alpha=0.2$). Here, we set the BH mass $M=1$ and constant parameter $\ell=0.05$.} \label{Fig2}
\end{figure} 
\begin{figure} [H]
	\begin{center}
    \begin{subfigure}[]
     {\includegraphics[width=0.55\textwidth,height=0.3\textheight]{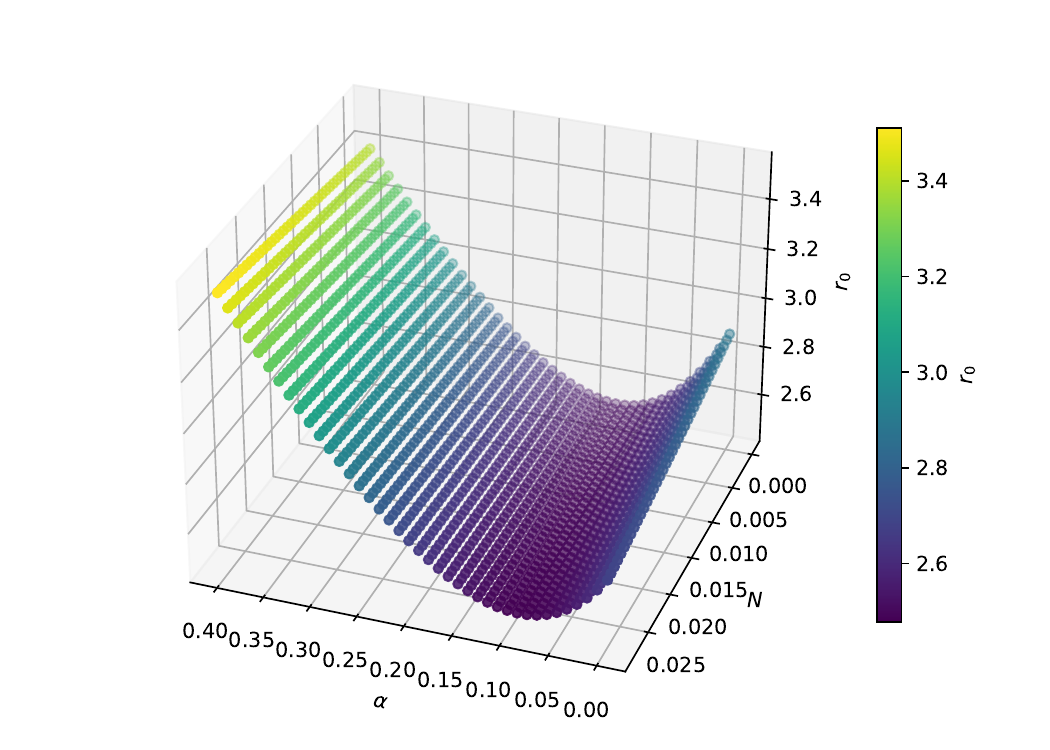}}   
    \end{subfigure}
     \begin{subfigure}[]
     {\includegraphics[width=0.4\textwidth,height=0.3\textheight]{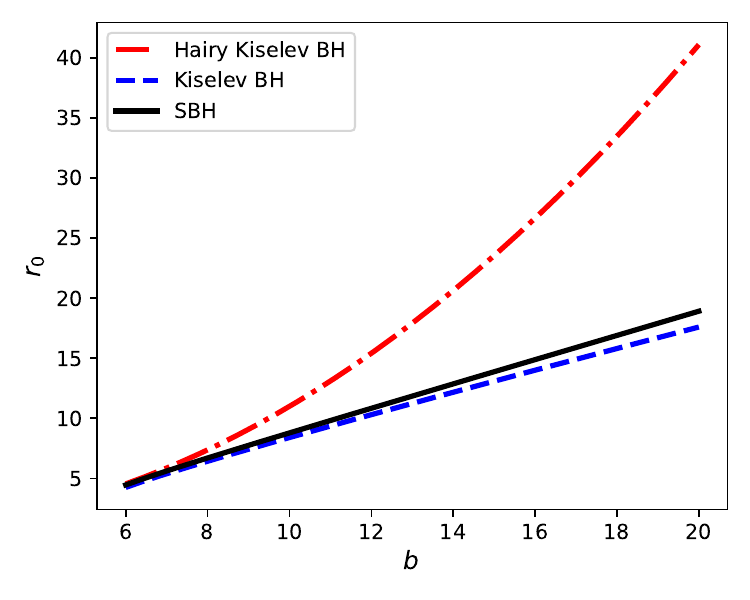}}  
    \end{subfigure}   
	\end{center}
	\caption{ (a) Three dimensional scatter plot of distance of closest approach as a function of $\alpha$ and $N$. (b) The comparison with other BH solution in which Hairy Kiselev BH reduces under specific limits. The sets of parameter for particular BHs we considered namely, Hairy Kiselev BH ($N=0.007, \alpha=0.2$), Kiselev BH ($N=0, \alpha=0.2$) and SBH ($N=0, \alpha=0$). } \label{Fig2.1}
\end{figure}
\noindent where, $\eta=\alpha e^{\frac{2r_0}{(2M-\alpha l)}} $. Thus, Eq. \ref{meomp} becomes,
\begin{equation} \label{BINU}
    B(u) = 2M (u-u_{1})(u-u_{2})(u-u_{3}). 
\end{equation} 
Substituting Eq. \ref{BINU} in \ref{eqopath} yields,
\begin{equation} \label{BINU2}
    \frac{du}{d\phi} = \pm \sqrt{2M (u-u_{1})(u-u_{2})(u-u_{3})}.
\end{equation}
The positive sign shows that the angle $\phi$; changes more than $\pi$, that is, the
photon trajectory is bent toward BH and for the negative sign the photon trajectory is bent
away from the BH. Now we can write the Eq. \ref{criticalimpact} in terms of distance of closest approach, 
\begin{equation}
    r_{0}^{3\omega+3} 
    + \alpha \gamma b^{2} r_{0}^{3\omega+2} 
    - \alpha b^{2} r_{0}^{3\omega+1} - b^{2} r_{0}^{3\omega+1} 
     + N b^{2} + 2 M b^{2} r_{0}^{3\omega} = 0.
\end{equation}
Here, $\gamma=\frac{2}{2M-\alpha l}$. Although an exact analytical solution to this equation is not readily attainable due to its complexity, it becomes tractable under certain conditions. Specifically, $\omega=-2/3$, the above equation reduces to a cubic form, which allows for an analytical solution. Considering the special case ($\omega=-2/3$), the equation becomes
\begin{equation}
    r_{0}^{3} +  \alpha \gamma  b^{2} r_{0}^{2} + N b^{2} r_{0}^{2} - \alpha b^{2}  r_{0} - b^{2} r_{0} + 2 M b^{2} = 0.
\end{equation}
Using Cardano’s method solving the cubic equation,
\begin{equation}
 r_{0}^{3} +  (\alpha \gamma b^{2} + N b^{2} ) r_{0}^{2} - (\alpha b^{2} + b^{2}) r_{0} + 2 M b^{2} = 0,
\end{equation}
the relation between $b$ and $r_{0}$  is,

\begin{align}
r_{0} &= 2 \sqrt{ \frac{3\left[(\alpha b^{2} + b^{2}) 
          + (\alpha \gamma b^{2} + N b^{2})^{2} \right]}{9} } \nonumber \\
         & \times \cos\Bigg[ \frac{1}{3} \arccos\Bigg( 
      \quad -\frac{2 (\alpha \gamma b^{2} + N b^{2})^{3} 
          + 9 (\alpha \gamma b^{2} + N b^{2})(\alpha b^{2} + b^{2}) 
          + 54 M b^{2}}{ \left[ 18(\alpha b^{2} + b^{2}) 
          + 6 (\alpha \gamma b^{2} + N b^{2}) \right] } \nonumber \\
      &\quad \times \sqrt{ \frac{9}{3\left[(\alpha b^{2} + b^{2}) 
          + (\alpha \gamma b^{2} + N b^{2})^{2} \right]} } \Bigg) 
          \Bigg] - \frac{(\alpha \gamma b^{2} + N b^{2})}{3}.
\end{align}
At $\alpha=0$, it consistently reduces to the Kiselev BH case \cite{Younas_2015},
\begin{equation}
r_{0(Kiselev)} = 2 \sqrt{\frac{\sigma^2 b^4 + 3b^2}{9}} \cos\left[ \frac{1}{3} \cos^{-1} \left(
\frac{-2\sigma^3 b^6 + 9\sigma b^4 + 54M b^2} {6\sigma^2 b^4 + 18b^2} \sqrt{\frac{9}{\sigma^2 b^4 + 3b^2}} \right) \right] - \frac{\sigma b^2}{3}.
\end{equation}
At $N=0$ and $\alpha=0$, it consistently reduces to the Schwarzschild BH case \cite{Iyer:2006cn},
\begin{equation}
r_{0(SBH)} = \frac{2b}{\sqrt{3}}  \cos\left[ \frac{1}{3} \cos^{-1} \left( -\frac{3\sqrt{3}M}{b} \right) \right].
\end{equation}
\noindent Fig.~\ref{Fig2} illustrates the variation of the distance of the closest approach as a function of the impact parameter for different values of the coupling constant $\alpha$ and the surrounding field parameter $N$ in the hairy Kiselev BH spacetime. It is evident that the distance of the closest approach increases with an increase in the impact parameter, as expected from general relativistic considerations. Furthermore, for fixed impact parameter values, the plots show that the closest approach distance attains its maximum for higher values of $\alpha$, indicating that a stronger coupling between the scalar field and gravity leads to a weaker effective gravitational pull. A similar behavior is observed with respect to $N$; the closest approach increases with increasing $N$, suggesting that the influence of the surrounding matter field also reduces the effective gravitational attraction. These observations highlight the significant role played by both the scalar coupling parameter $\alpha$ and the surrounding field parameter $N$ in the governing  of photon trajectories and gravitational lensing in such modified BH geometries, shown in subfigure~(a) of Fig.~\ref{Fig2.1}. In subfigure~(b) of Fig.~\ref{Fig2.1}, we compare the behavior for the Hairy Kiselev BH, the standard Kiselev BH, and the SBH. The results show that the Hairy Kiselev BH allows the largest closest approach distance, followed by the SBH, while the standard Kiselev BH exhibits the smallest. This indicates that the presence of scalar hair in the Hairy Kiselev BH modifies the effective geometry more strongly than in the other two cases, thereby influencing photon trajectories significantly.

\section{Bending Angle} \label{Sec5}
\begin{figure} [H]
	\begin{center}
    \begin{subfigure}[]
     {\includegraphics[width=0.45\textwidth]{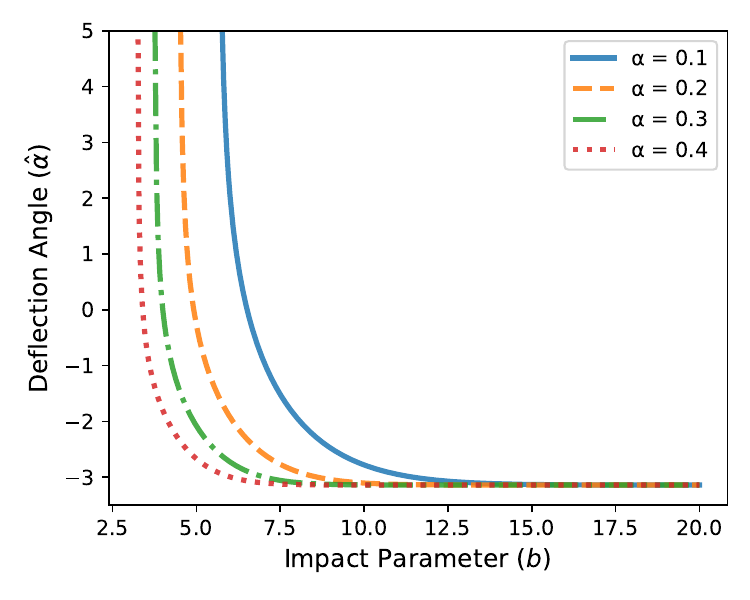}}   
    \end{subfigure}
     \begin{subfigure}[]
     {\includegraphics[width=0.45\textwidth]{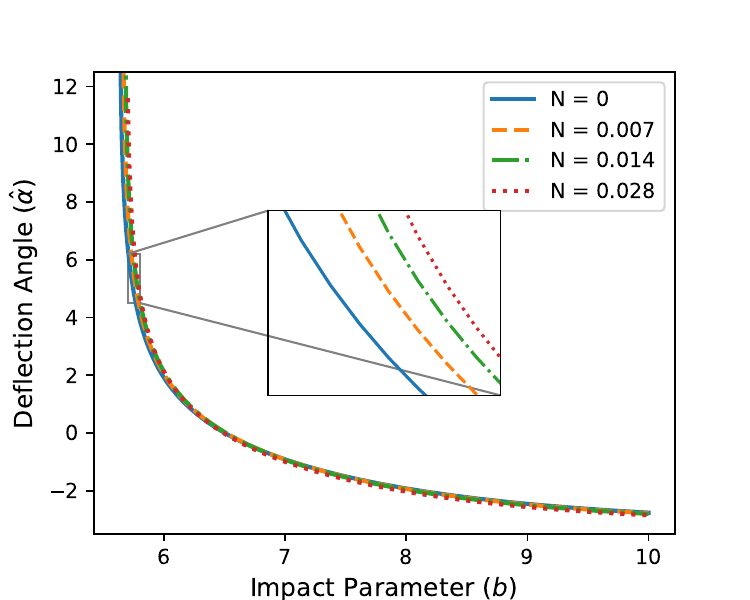}}  
    \end{subfigure}   
	\end{center}
	\caption{Bending angle as a function of impact parameter is shown for varying values of the coupling parameter $\alpha$ (left panel, $N=0.007$) and the normalization constant $N$ of the field (right panel, $\alpha=0.2$). Here, we set the BH mass $M=1$ and constant parameter $\ell=0.05$.} \label{Fig3}
\end{figure}

\begin{figure} [H]
	\begin{center}
    \begin{subfigure}[]
     {\includegraphics[width=0.45\textwidth]{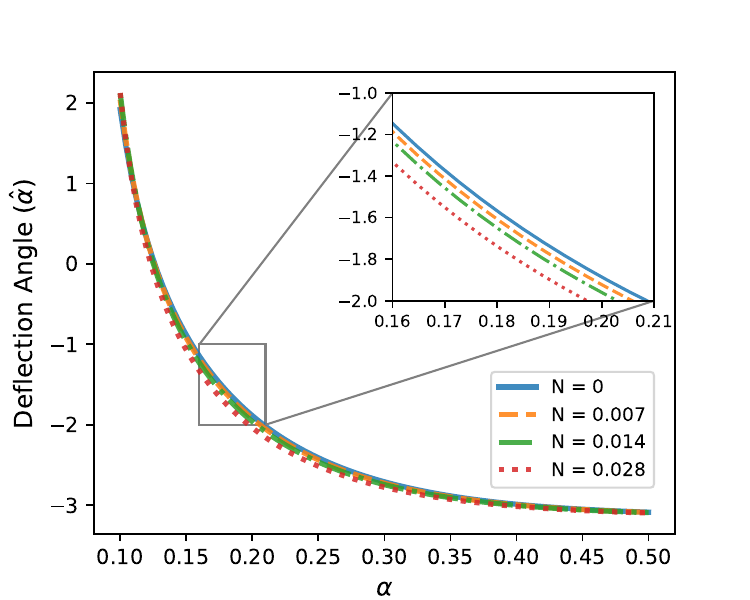}}   
    \end{subfigure}
     \begin{subfigure}[]
     {\includegraphics[width=0.45\textwidth]{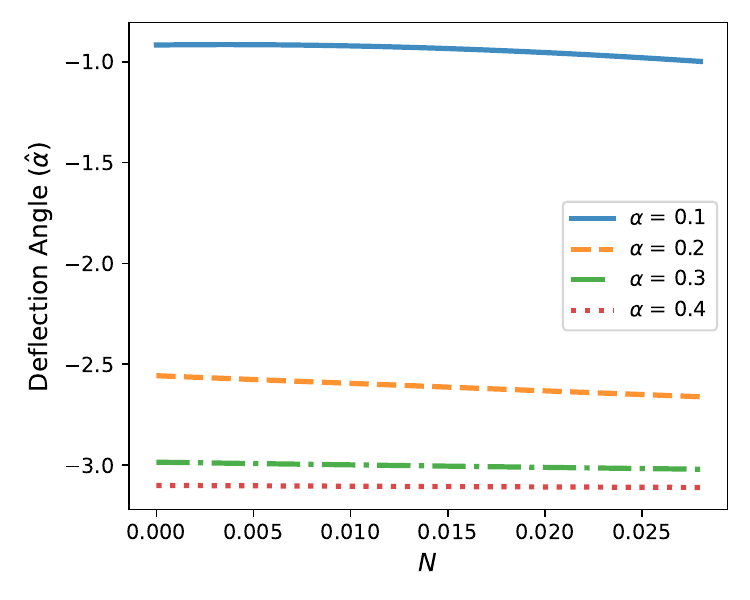}}  
    \end{subfigure}   
	\end{center}
	\caption{Bending angle as a function of coupling constant and surrounding field parameter. Here, we set the BH mass $M=1$ and constant parameter $\ell=0.05$.} \label{Fig4}
\end{figure}

\noindent In this section, we investigate the gravitational deflection of light in the vicinity of a BH. Specifically, we consider the scenario in which a light ray originates from spatial infinity, approaches the BH to a minimum radial distance $r_0$ referred to as the point of closest approach and subsequently recedes back to infinity, where it is detected by a distant observer. During this journey, the angular coordinate $\phi$ undergoes a total change due to the curvature of spacetime induced by the BH. This results in a deviation of the light path from its original straight-line trajectory. The total bending of the light ray, known as the deflection angle $\hat{\alpha}$, is quantified by
\begin{equation}
    \hat{\alpha} = 2 \int_{0}^{1/r_0} \frac{d\phi}{du} \, du - \pi
\end{equation}
If we substitute Eq. \ref{BINU2} into this, we get:
\begin{equation}
    \hat{\alpha} = 2 \int_{0}^{1/r_0} \frac{1}{\sqrt{2M(u - u_1)(u - u_2)(u - u_3)}} \, du - \pi
\end{equation}
The integral can be recast in terms of elliptic integrals by decomposing the integration range at the radial turning points of the photon trajectory. This yields the following expression,
\begin{align}
    \hat{\alpha} = \frac{\sqrt{2}}{M} \Bigg[ 
        & \int_{u_1}^{1/r_0} \frac{1}{\sqrt{(u_1 - u)(u - u_2)(u_3 - u)}} \, du \\
        & - \int_{0}^{u_1} \frac{1}{\sqrt{(u_1 - u)(u - u_2)(u_3 - u)}} \, du 
    \Bigg] - \pi
\end{align}
To facilitate analytical evaluation, the integral is reformulated in terms of elliptic integrals of the first kind, assuming an ordering of the turning points such that (\( u_3 > u_2 > u_1 \)). This transformation allows us to express the result using standard elliptic functions as follows.
\begin{equation}
    \hat{\alpha} = \frac{2\sqrt{2}}{M} \left[ \frac{F(\Psi_1, k)}{\sqrt{u_3 - u_1}} - \frac{F(\Psi_2, k)}{\sqrt{u_3 - u_1}} \right] - \pi.
\end{equation}
Here, $F(\Psi, k)$ is the incomplete elliptic integral of the first kind, which is defined as
\begin{equation}
    F(\Psi, k) = \int_0^{\Psi} \frac{d\theta}{\sqrt{1 - k^2 \sin^2 \theta}}.
\end{equation}
For a detailed explanation of elliptic integrals, we refer the reader to the book by Byrd et al. \cite{byrd2013handbook}.
The integral variables appearing in the expression for the bending angle are defined as follows,
    \begin{align}
    \Psi_1 &= \frac{\pi}{2}, \\
    \Psi_2 &= \sin^{-1} \left( 
        \sqrt{ \frac{
            r_0 - 2M - N r_0 - \eta r_0 -  r_0 \left(\sqrt{\left(\frac{2M}{r_{0}}+N+\eta-1\right)^{2}-8M \left(\frac{2M+\eta}{r_{0}^{2}}+\frac{N-1}{r_{0}}\right)}\right)
        }{
            r_0 - 6M - N r_0 - \eta r_0 -  r_0 \left(\sqrt{\left(\frac{2M}{r_{0}}+N+\eta-1\right)^{2}-8M \left(\frac{2M+\eta}{r_{0}^{2}}+\frac{N-1}{r_{0}}\right)}\right)
        }}
    \right).
\end{align}
In the context of elliptic integrals, the elliptical modulus \(k\) is defined such that \(0 \leq k^2 \leq 1\). It is worth noting that some references denote this variable as \(k^2\), while others use \(k\) to represent the modulus. In our case, it is given as,
    \begin{equation}
    k = \frac{
        \sqrt{6M - r_0 +\sqrt{\left(\frac{2M}{r_{0}}+N+\eta-1\right)^{2}-8M \left(\frac{2M+\eta}{r_{0}^{2}}+\frac{N-1}{r_{0}}\right)}}
    }{
        2\sqrt{\left(\frac{2M}{r_{0}}+N+\eta-1\right)^{2}-8M \left(\frac{2M+\eta}{r_{0}^{2}}+\frac{N-1}{r_{0}}\right)}
    }.
\end{equation}
Thus, an exact expression for the bending angle can be obtained:
    \begin{equation}
    \hat{\alpha} = 4 \sqrt{\frac{r_{0}}{\sqrt{\left(\frac{2M}{r_{0}}+N+\eta-1\right)^{2}-8M \left(\frac{2M+\eta}{r_{0}^{2}}+\frac{N-1}{r_{0}}\right)}}} \left[ K(k) - F(\Psi, k) \right] - \pi.
\end{equation}
In the limiting case \(\alpha = 0\), the derived expression for the bending angle reduces to the result for the Kiselev BH, as obtained in \cite{Younas_2015}. Moreover, in the simultaneous limit \(N = 0\) and \(\alpha = 0\), it further simplifies the standard Schwarzschild BH case, consistent with the findings of \cite{Iyer:2006cn}.
The graphical representation of the bending (deflection) angle as a function of the impact parameter for various values of the coupling constant $(\alpha)$ and the surrounding field parameter $(N)$ is shown in Fig.~\ref{Fig3}. As a fundamental feature of gravitational lensing, we observe that the deflection angle decreases monotonically with increasing impact parameter. This behavior aligns with the physical expectation that light rays passing closer to the BH experience stron-ger spacetime curvature and hence greater deflection. Additionally, our analysis reveals that the deflection angle is more pronounced for lower values of the coupling constant. However, as $\alpha$ continues to increase, the rate of increase in the deflection angle diminishes, suggesting a saturating effect of the scalar field coupling on the gravitational interaction. This reflects how the scalar field, mediated via the coupling constant, enhances the effective gravitational field, particularly in the strong field regime.
 Fig.~\ref{Fig4} illustrates the variation of the deflection angle as a function of the coupling constant \(\alpha\) and the surrounding field parameter \(N\). \noindent In the subfig $(a)$ of Fig.~\ref{Fig4}, it is observed that the deflection angle decreases with increasing \(\alpha\), indicating that stronger coupling suppresses the light bending. Although multiple values of \(N\) are considered, its effect appears to be subdominant compared to \(\alpha\), as noted previously. To examine the subtle influence of \(N\), a zoomed-in region is included, which reveals that the deflection angle is marginally higher for smaller values of \(N\), suggesting that a weaker quintessence-like field enhances the gravitational lensing. Further, the subfig $(b)$ of Fig.~\ref{Fig4} shows a mild decrease in the deflection angle with increasing \(N\), and the maximum deflection occurs for the smallest \(\alpha\). The emergence of negative deflection angles in certain regimes implies a repulsive behavior or nontrivial photon trajectory bending, possibly due to the combined effects of the scalar hair and the surrounding field. This behavior is consistent with theoretical expectations for configurations involving exotic matter distributions or modifications to standard GR. 

\begin{figure} [H]
	\begin{center}
    \begin{subfigure}[]
     {\includegraphics[width=0.45\textwidth,height=0.3\textheight]{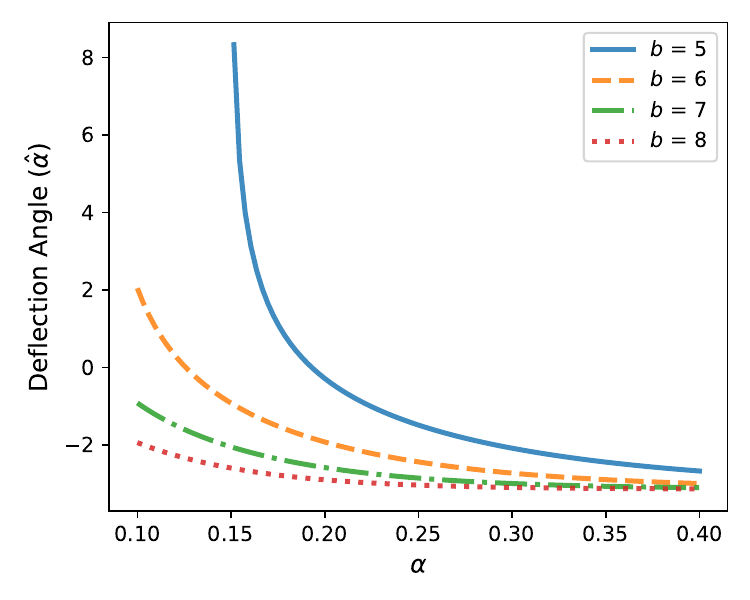}}   
    \end{subfigure}
     \begin{subfigure}[]
     {\includegraphics[width=0.45\textwidth,height=0.3\textheight]{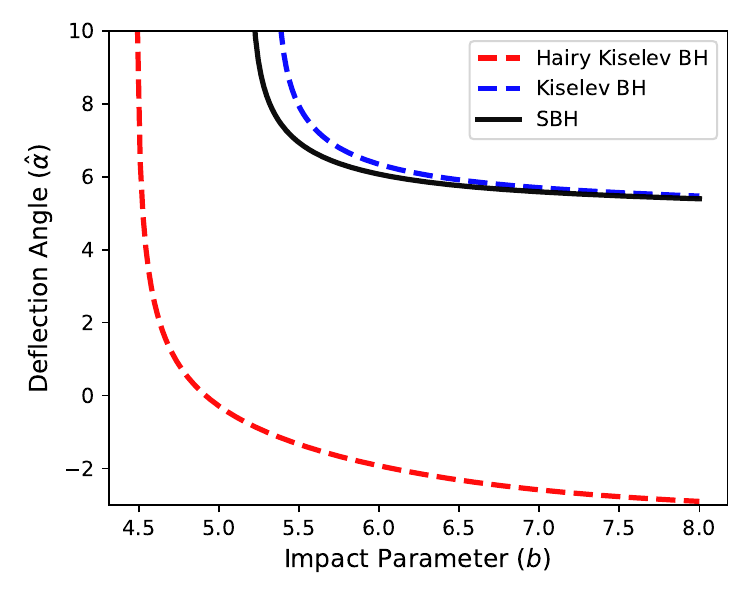}}  
    \end{subfigure}   
	\end{center}
	\caption{(a) Bending angle as a function of coupling constant for different values of impact parameter. (b) The comparison of bending angle with Kiselev BH and SBH. The sets of parameter for particular BHs we considered namely, Hairy Kiselev BH ($N=0.007, \alpha=0.2$), Kiselev BH ($N=0, \alpha=0.2$) and SBH ($N=0, \alpha=0$). } \label{Fig5}
\end{figure}
\vspace{1cm}
The subfig $(a)$ of of Fig.~\ref{Fig5} illustrates the variation of the deflection angle with respect to the coupling constant for different values of the impact parameter. It is evident that as the impact parameter increases, the deflection angle decreases. This behavior is consistent with GL theory, where light rays passing closer to the BH (i.e., smaller impact parameters) experience stronger spacetime curvature, resulting in larger deflection angles. Furthermore, the subfig $(b)$ of of Fig.~\ref{Fig5} presents a comparative analysis between the Hairy Kiselev BH, the Hairy BH, and the Schwarzschild BH (SBH). The results indicate that the deflection angle for the Kiselev BH is greater than that for the SBH, primarily due to the additional energy density contributed by the surrounding quintessential field. Conversely, the deflection angle for the Hairy Kiselev BH is significantly smaller than that of the SBH. This reduction can be attributed to the presence of the scalar hair, which effectively modifies the spacetime geometry and reduces the gravitational lensing strength. The findings thus clearly demonstrate that the introduction of the coupling (hairy) parameter leads to a suppression of the deflection angle, highlighting the impact of scalar fields on light propagation in modified gravity scenarios.


\section{CONCLUSIONS} \label{Sec6}
\noindent
In summary, the bending of light rays around a Hairy Kiselev BH is investigated, with particular emphasis on the effects of the scalar field coupling constant and the quintessence parameter on the deflection angle, specifically for the case $\omega = -\frac{2}{3}$. We begin by analyzing the BH horizon structure and establish the crucial role of the state parameter $(\omega)$ in determining the causal structure of spacetime. For $\omega = -\frac{4}{3}$ and $\omega = -1$, the metric admits two distinct horizons. In contrast, for $\omega = -\frac{2}{3}$ and $\omega = 0$, only a single event horizon is formed, highlighting the sensitivity of horizon formation to the nature of the surrounding field.
Further, we derived the geodesic equations and analyzed the corresponding effective potential, which revealed the existence of unstable circular photon orbits.
The critical parameters were then computed, and the relationship between the closest approach distance and the impact parameter was examined. The analysis of the photon sphere clearly suggests that its radius increases with increasing values of both parameters. This behavior indicates that the presence of the surrounding field and scalar coupling enhances the light deflection region, thereby extending the unstable photon orbit farther from the BH. Finally, we obtained an exact analytical expression for the deflection angle in terms of elliptic integrals, allowing for a precise characterization of light bending in this modified gravity background. Our analysis demonstrates that the deflection angle decreases with increasing impact parameter, consistent with classical gravitational lensing predictions. The presence of a scalar field, governed by the coupling constant (\(\alpha\)), significantly modifies the deflection behavior: while a higher \(\alpha\) initially enhances the bending, it ultimately leads to a suppression in the strong-field regime due to spacetime modifications. In the limiting case \(\alpha = 0\), the derived expression for the bending angle reduces to the result for the Kiselev BH, as obtained in \cite{Younas_2015}. Moreover, in the simultaneous limit \(N = 0\) and \(\alpha = 0\), it further simplifies the standard Schwarzschild BH case, consistent with the findings of \cite{Iyer:2006cn}. Comparisons among Schwarzschild, Kiselev, and Hairy Kiselev BHs reveal that the quintessential field (parameter \(N\)) enhances deflection, whereas scalar hair reduces it. Importantly, our analysis reveals qualitatively distinct behavior across the weak-field and strong-field regimes. In the weak-field limit, the deflection angle asymptotically recovers the classical expression derived by Iyer et al. \cite{Iyer:2006cn}, demonstrating consistency with general relativistic predictions far from the photon sphere. In contrast, in the strong-field regime near the photon sphere, the presence of the scalar coupling constant introduces significant deviations that are absent in the limiting cases where the Hairy Kiselev BH reduces. This deviation reflects the nonlinear influence of scalar hair and the quintessential field, particularly in high-curvature regions, leading to a measurable shift in the photon sphere radius and an alteration of the deflection profile. Such modifications offer promising theoretical avenues to probe scalar-tensor extensions of GR through precise observations of strong GL phenomena, including photon rings and shadow boundaries around supermassive BHs. These theoretical trends are compatible with the EHT observations of M87* and Sgr A*, which already constrain deviations from GR, while the next-generation EHT Collaboration promises to open new windows to test and possibly confirm these theoretical predictions.
\newpage
\section{Acknowledgments}
The author, SK, sincerely acknowledges IMSc for providing exceptional research facilities and a conducive environment that facilitated his work as an Institute Postdoctoral Fellow. The authors, HN and FA would like to thank IUCAA, Pune, for the support under its associateship program, where a part of this work was done. HN also acknowledges the financial support provided by the Science and Engineering Research Board (SERB), New Delhi, through grant number CRG/2023/-008980. SR express sincere and deep gratitude to the Department of Physics, NITA, for providing the necessary research environment to complete this work.
\bibliography{main}
\bibliographystyle{unsrt}
\end{document}